\documentclass[letter,twocolumn,showpacs,preprintnumbers,prl,byrevtex]{revtex4-1}

\usepackage{graphics}% Include figure files
\usepackage{amsmath}
\usepackage{amssymb}
\usepackage{bm}% bold math
\usepackage{graphicx}% Include figure files, able to rotate

\begin{document}
\title{Phase Space Formulation of Population Dynamics in Ecology}
\author{Jes\'us Mart\'{\i}nez-Linares}
\address{Departamento de F\'{i}sica Aplicada II.Universidad de Sevilla.  \\
41012-Seville, Spain. }

%\date{Received \phantom{31 December 1999}}
%\date{\today}
\date{May 26, 2007}

\begin{abstract}
A phase space theory for population dynamics in Ecology is
presented. This theory applies for a certain class of dynamical
systems, that will be called $\mathcal{M}$-systems,  for which a
conserved quantity, the $\mathcal{M}$-function, can be defined in
phase space. This $\mathcal{M}$-function is the generator of time
displacements
%can be interpreted as a measure of a
%mutual biological resource at disposal of the interacting species
and contains all the dynamical information of the system. In this
sense the $\mathcal{M}$-function plays the role of the hamiltonian
function for mechanical systems.
%In fact, as the latter, it can be
%regarded as the generator of time displacements.
In analogy with Hamilton theory we derive equations of motion as
derivatives over the resource function in phase space.  A
$\mathcal{M}$-bracket is defined which allows one to perform a
geometrical approach in analogy to Poisson bracket of hamiltonian
systems. We show that the equations of motion can be derived from a
variational principle over a functional $\mathcal{J}$ of the
trajectories. This functional plays for $\mathcal{M}$-systems the
same role than the action $S$  for hamiltonian systems. Finally,
three important systems in population dynamics, namely,
Lotka-Volterra, self-feeding and logistic evolution, are shown to be
$\mathcal{M}$-systems.
\end{abstract}

%87.23.Cc Population dynamics and ecological pattern formation
%87.10.+e General theory and mathematical aspects
%87.23.-n Ecology and evolution

\narrowtext

\maketitle

%
%%%%%%%%%%%%%%%%%%%%%%%%%%%%%%%%%%%%%%%%%%%%%%%%%%%%%%%%%%%%%%%%%%%%%%%%%%%%%%%%
%%%%%%%%%%%%%%%%%%%%%%%%%%%%%%%%%%%%%%%%%%%%%%%%%%%%%%%%%%%%%%%%%%%%%%%%%%%%%%%%
%

%\section{Introduction}
%
%%%%%%%%%%%%%%%%%%%%%%%%%%%%%%%%%%%%%%%%%%%%%%%%%%%%%%%%%%%%%%%%%%%%%%%%%
%

Conservation of energy modeled by differential geometry is the
cornerstone of the phase space formulation of classical mechanics
\cite{Goldstein}. There exist a wide machinery of analytical methods
developed during the past two centuries for such systems
\cite{Saletan}. For instance, the Hamilton formalism resorts on a
Hamiltonian function defined on phase space which contains all the
dynamical information of the system. Once specified, we can generate
upon differentiation equations of motion for the system.

The situation is much more precarious in population dynamics in
Ecology. Ecological systems are overwhelmingly complex systems
\cite{Begon}. However, in some cases they can be described by
simplified phenomenological rate equations that describes the net
effect of the environment and the interaction between densities of
species \cite{Hoppensteadt}.
%The net
%effect of the environment and natural birth and death rates can then
%be also incorporated in the equations \cite{Hoppensteadt}.
However, a systematic approach allowing one to derive equations of
motion from general principles is still lacking. The present paper
is a step towards this goal. In fact,
%Individuals need resources to live. The evolution can be described
%as a redistribution of resources at a given time scale. Usually, the
%phenomenological equations of motion lose track of the sources and
%drains of resources. However,
in some cases a mutual resource function  is conserved and keep
constant during the evolution  of the interacting species. The
question is then, can be develop a phase space formalism that
expresses the conservation of this quantity?

We show in this paper that there exist a wide class of systems in
ecology, that will be call $\mathcal{M}$-systems, for which
population dynamics can be described by a phase space theory
analogous to the hamiltonian formulation of classical mechanics. We
exploit this analogy in order to use the elaborated analytical
methods developed for the latter. Hamilton function, Poisson
bracket, variational principles, action functional and Lagrangian,
all of them useful tools for hamiltonian systems will find their
counterparts in the theory.

%For instance, we show that the equation of motion of
%$\mathcal{M}$-systems can be derived from a variational principle
%over the trajectories in phase space. This allows us to find a
%functional $\mathcal{J}$ with the following property: in
%$\mathcal{M}$-systems, the actual trajectories are such that make
%$\mathcal{J}$ an extremal ($\delta \mathcal{J}=0$).
Finally, we show
that three important system in population ecology are
$\mathcal{M}$-systems. They are the Lotka-Volterra predator-prey
system, the self-feeding interaction and the logistic evolution.
These systems are also applied in many other disciplines outside
ecosystems, which sets forth the wide range of applications of the
$\mathcal{M}$-system formalism.

We start with the definition of $\mathcal{M}$-systems. 
%On the other hand, the main goal of population dynamics is to
%describe the evolution of the population of different species
%interacting in an ecological system.
%When $\alpha$ is even
%\cite{commentparity}, it is possible to choose a phase space to be
%the carrier manifold of the population dynamics.
A $\mathcal{M}$-system is  a dynamical system defined by a given
resource function  $\mathcal{M}(q,p\: ;t)$, and
equations of motions given by
\begin{eqnarray}
\dot{q}&=& qp\; \frac{\partial \mathcal{M}}{\partial p} ,
\nonumber\\
\dot{p}&=& -qp\; \frac{\partial \mathcal{M}}{\partial q}.
\label{MEM}
\end{eqnarray}
%These equations are highly non-linear. Note that in contrast with
%Eqs. (\ref{HEMM}), Eqs. (\ref{MEM}) yield non-linear equations of
%motion for lineal $\mathcal{M}$-functions.

Not every conceivable motion in phase space  $\Gamma=(q,p)$ is a
$\mathcal{M}$-dynamical system. In order to obtain a necessary condition let
us consider the generic dynamical system
\begin{eqnarray}
\dot{q}&=& g_1(q,p) ,
\nonumber\\
\dot{p}&=& g_2(q,p) ,\label{generic}
\end{eqnarray}
where $g_1, g_2 \in \mathcal{F}(\Gamma%\mathbf{T}^*\mathbb{Q}
)$ are
two arbitrary functions in phase space. The existence of a
continuous $\mathcal{M}$-function requires
\begin{equation}
 \frac{\partial^2 \mathcal{M}}{\partial q \partial p}= \frac{\partial^2 \mathcal{M}}{\partial p \partial
 q}.
\label{cruzado}
\end{equation}
Eq. (\ref{cruzado}), together with (\ref{MEM}) and (\ref{generic})
yields
\begin{equation}
 \frac{g_1}{q}+ \frac{g_2}{p}= \frac{\partial g_1}{\partial q}+ \frac{\partial g_2}{\partial
 p},
 \label{condicion}
\end{equation}
which is a necessary condition for the dynamical system
(\ref{generic}) to be a $\mathcal{M}$-system. If this is the case we
say that $g_1, g_2$ are the components of a $\mathcal{M}$-vector
field.

According to Eq. (\ref{MEM}), the resource function  $\mathcal{M}$ characterizes the dynamics of the system. The $\mathcal{M}$-systems are not hamiltonian systems, since they do not obey the hamiltonian canonical equations of motion. However, they possess a close relationship with each other. In fact, consider the change of variables
\begin{eqnarray}
q' &=& \ln (q/\sigma),
\nonumber\\
p\,' &=& \ln (p/\xi),
 \label{changeofvariable}
\end{eqnarray}
where $\sigma$, $\xi$ are two positive constants. Using Eqs. (\ref{MEM}) it is easy to check that they obey the hamilton canonical equations \cite{Goldstein}
%First, recall
%that the Hamilton's canonical equations of motion in phase space are
%given by
\begin{eqnarray}
\dot{q}'&=& \; \frac{\partial \mathcal{H}}{\partial p\,'} ,
\nonumber\\
\dot{p}\,'&=& -\; \frac{\partial \mathcal{H}}{\partial q'}, \label{HEMM}
\end{eqnarray}
where $\mathcal{H}(q',p')=\mathcal{M}(q(q'),p(p\,'))$ is the hamiltonian function. 
Thus,  Eqs. (\ref{changeofvariable}) allows one to construct an associated  $\mathcal{M}$-system starting from a generic hamiltonian system. Conversely, note that these relations comprise the Hamilton phase space $\Lambda=\mathbb{R}^{2}$ into the first octant $\mathbb{R}^{2}_+=(q,p\in \mathbb{R}, q,p\geq 0)$. Thus, the class of $\mathcal{M}$-system on $\Gamma=\mathbb{R}^{2}_+$ has an associated hamiltonian system on $\Lambda$. We will discuss later the importance of this system sub-class in ecology and game theory.

For many dynamical systems $\mathcal{H}$ gives their mechanical
energy E. The resource function $\mathcal{M}$ play for $\mathcal{M}$-systems
the same role than the function $\mathcal{H}$ for hamiltonian
systems. In order to see this, we introduce a geometrical approach
analogous to the Poisson bracket of hamiltonian system.
%\section{The  $\mathcal{M}$-bracket}
%\label{sectionMB}
The Poisson bracket (PB) %plays a mayor role in hamiltonian systems.It
is defined as \cite{Goldstein}
\begin{equation}
 \{f,g\}_{\text{PB}}=(\partial_j f)\; \Omega^{jk} (\partial_k g)
\label{PB}
\end{equation}
where $f,g \in \mathcal{F}(\Gamma%\mathbf{T}^*\mathbb{Q}
)$ are two
arbitrary functions on phase space, $\Omega$ is the symplectic
matrix
\begin{equation}
\label{OB}
 \Omega =  \left(
\begin{array}{cc}
0 & 1 \\
-1 & 0
\end{array}
 \right),
\end{equation}
and we have defined the derivative operators $\partial_i \equiv
\frac{\partial}{\partial \eta_i}$ where $\eta_1=q$, $\eta_2=p$. We
follow the Einstein rule of summation over repeated indexes. In
terms of the PB, Eqs. (\ref{HEMM}) can be written as
\begin{equation}
 \dot{\eta_i}=\{\eta_i,\mathcal{H}\}_{\text{PB}}.
\label{HEME}
\end{equation}
Moreover, the evolution of any dynamical variable $f(q,p\: ;t) \in
\Gamma%\mathbf{T}^*\mathbb{Q}
$ is given by
\begin{equation}
 \frac{df}{dt}=\{f,\mathcal{H}\}_{\text{PB}} +\frac{\partial f}{\partial t}.
\label{fPB}
\end{equation}
 According to (\ref{fPB}), $\mathcal{H}$ is the generator of time
displacements of the system. Due to the antisymmetry of the PB, the
time evolution of the hamiltonian is
$\dot{\mathcal{H}}=\frac{\partial \mathcal{H}}{\partial t}$. This
reflects the conservation of energy for time-independent hamiltonian
systems.

We now define the $\mathcal{M}$-bracket (MB) as
\begin{equation}
 [f,g]_{\text{MB}}=(\widetilde{\partial}_j f) \;\Omega^{jk} (\widetilde{\partial}_k
 g),
\label{MB}
\end{equation}
where we have defined the derivative operators
$\widetilde{\partial}_i \equiv \frac{\partial}{\partial
\ln(\eta_i)}$. It can be shown that MB (as the PB) is bilinear,
antisymmetric and satisfies the Jacobi identity.
%, as shown in Appendix \ref{AppendixI}.
These are the defining properties of a Lie algebra \cite{Sat86}. The
function space $\mathcal{F}(\Gamma%\mathbf{T}^*\mathbb{Q}
)$ is a Lie
algebra under the action of the MB. MB also satisfies the product
rule
\begin{equation}
 [f,g h]_{\text{MB}}=[f,g]_{\text{MB}}h+g[f,h]_{\text{MB}}.
\label{productrule}
\end{equation}
Thus, MB is a kind of derivative operator \cite{Moyal}. As a matter
of fact, Eqs. (\ref{MEM}) can be recast in terms of the MB as
\begin{equation}
 \dot{\eta_i}=[\eta_i,\mathcal{M}]_{\text{MB}}.
\label{HEM}
\end{equation}

The function $\mathcal{M}\in
\mathcal{F}(\Gamma%\mathbf{T}^*\mathbb{Q}
)$ is the generator of time
displacements on the system, since the time derivative of any
dynamical variable $f$ along the motion is
\begin{equation}
 \frac{df}{dt}=[f,\mathcal{M}]_{\text{MB}}+\frac{\partial f}{\partial t}.
\label{dot}
\end{equation}
In particular, choosing $\mathcal{M}=f$ we obtain
$\dot{\mathcal{M}}=\frac{\partial \mathcal{M}}{\partial t}$, i.e.,
the resource function of the system is conserved if it does not
depend explicitly on time. The equation $\mathcal{M}(q,p)=R$ for
$\mathcal{M}$-systems plays the same role than the equation
$\mathcal{H}(q,p)=E$ for hamiltonian systems.

%% OJO%%%%%%%%%%%%%%%%%%%%%%%%%%%%%%%%%%%%%%%%%%%%%%%%%%%%%%%%%%%%%%
%%It can be shown that a necessary and sufficient condition for the
%%existence of a $mathcal{M}$-vector field by transposing the
%%derivation for Hamiltonian systems given in pg. 223 of Saletan.

%\section{Variational Principle}
%\label{action}
Hamiltonian systems admit a variational formulation. In fact,
dynamical systems in classical mechanics minimize the action
functional \cite{Goldstein}
\begin{equation}
 S=\int_{t_o}^{t_1} L(q'(t),\dot{q}'(t),t)\, dt
\label{S}
\end{equation}
where $L$ is the Lagrangian function of the system
\begin{equation}
 L=p\,'\,\dot{q}' -\mathcal{H}
\label{L}
\end{equation}
It can be shown \cite{Goldstein} that Hamilton canonical equations
given in (\ref{HEMM}) can be derived from the variational principle
$\delta S=0$. 
 With the help of Eqs. (\ref{changeofvariable}), we can now construct a variational principle for the associated $\mathcal{M}$-system. To this end we define the $\mathcal{M}$-Lagrangian
\begin{equation}
 \mathcal{L}=\frac{\dot{q}}{q} \ln (p/\xi) -\mathcal{M} ,
\label{LL}
\end{equation}
and the  $\mathcal{M}$-Action 
\begin{equation}
\mathcal{J}=\int_{t_o}^{t_1} \!\mathcal{L}\, dt.
\label{J}
\end{equation}
 It is easy to check \cite{commenthamiltonprinciple} that the imposition of the extreme condition to the $\mathcal{M}$-Action, i.e., $\delta \mathcal{J}=0$, renders   the canonical equations of motion for $\mathcal{M}$-systems given in Eqs. (\ref{MEM}). Thus, among all the possible trajectories that a  $\mathcal{M}$-system could undertake, the actual motion
$(q(t),p(t))$ is the one that makes Eq. (\ref{J}) minimal, i.e., $\delta
\mathcal{J}=0$, where $\delta \mathcal{J}$ is the variation of $\mathcal{J}$ over different
trajectories connecting $(q(t_o),p(t_o))$ and $(q(t_1),p(t_1))$.

%It can be rapidly checked that the imposition of Eqs. (\ref{dI}) to
%$\mathcal{L}$ renders the canonical equations of motion for
%$\mathcal{M}$-systems given in Eqs. (\ref{MEM}). Thus, we have found
%a general variational principle for $\mathcal{M}$-systems. Their
%actual trajectories make the functional
%\begin{equation}
% \mathcal{J}=\int_{t_o}^{t_1} \!\mathcal{L}\, dt
%\label{SJ}
%\end{equation}
%an extreme, i.e. $\delta \mathcal{J}=0$. 
 $\mathcal{J}$ and
$\mathcal{L}$ play for the $\mathcal{M}$-systems the role that the
mechanical action $S$ and the Lagrangian $L$ play for hamiltonian
systems.
The definition of $\mathcal{L}$ retains a certain amount of
ambiguity \cite{commentHMP}. As a matter of fact, we can construct
$\mathcal{L}$ in the form
\begin{equation}
 \mathcal{L}=\dot{q}\,\alpha+\dot{p}\,\beta -\mathcal{M},
\label{Lab}
\end{equation}
where $\alpha(q,p)$, $\beta(q,p)$ are generic functions in phase
space (dynamical variables) satisfying
\begin{equation}
 \frac{\partial \beta}{\partial q}= -\frac{1}{qp}+\frac{\partial\alpha}{\partial p}.
\label{ab}
\end{equation}

%\section{Examples of $\mathcal{M}$-systems}
%\label{examples}
%\subsection{The Lotka-Volterra model}
%\label{LV}

The main goal of population dynamics is to
describe the evolution of the population of different species
interacting in an ecological system. We show now that $\mathcal{M}$-systems comprises a wide class
of important system in ecology. Let us start with the Lotka-Volterra
(LV) model \cite{Lotka,Volterra} describing an ecological
predator-prey (or parasite-host) system.
%is one of the earliest predator-prey models to be based on sound mathematical
%principles.
%It forms the basis of many models used today in the analysis of
%population dynamics. For two-competing species they come from the
%pioneer work of Lotka on parasitology \cite{Lotka} and Volterra on
%fishing activity in the upper Adriatic sea \cite{Volterra}. The LV
%two-species model
%describe an ecological predator-prey (or parasite-host) system. It
The LV two-species model is given by the equations
\begin{eqnarray}
\frac{dU}{dt}&=&\left( a_1-b_1 V\right) U,
\nonumber\\
\frac{dV}{dt}&=&\left( -a_2+b_2 U\right) V, \label{O1}
\end{eqnarray}
where $a_1$ is the the growth rate of prey $U$, $b_1$ the rate at
which predators $V$ destroy the prey, $a_2$ the the death rate of
predators and $b_2$ the rate at which predators increase by
consuming prey. We take the adimensional variables proposed by Hsu
\cite{Hsu}
\begin{equation}
q=\frac{b_2}{a_2} \; U, \hspace{0.5cm} p=\frac{b_1}{a_2} \; V,
\hspace{0.5cm} \tau=a_2 t, \hspace{0.5cm} \xi=\frac{a_1}{a_2},
\end{equation}
to reduce the Lotka-Volterra (LV) system to
\begin{eqnarray}
\frac{dq}{d\tau}&=&\left( \xi-p\right) q,
\nonumber\\
\frac{dp}{d\tau}&=&\left( q-1\right) p.
 \label{O2}
\end{eqnarray}
%to be solved with the initial conditions $q(0)=q_o>0$ and
%$p(0)=p_o>0$.

%% In matrix form
%%\begin{equation}
%%\frac{d\vec{r}}{d\tau}=A\,\vec{r}+\vec{r}^T \Omega \;\vec{r},
%%\end{equation}
%%where we have defined the vector $\vec{r}=(u,v)$ and A is the drift
%%matrix
%%\begin{equation} \label{OA}
%% A = \left(
%%\begin{array}{cc}
%%\xi & 0 \\
%%0 & -1
%%\end{array}
%% \right),
%%\end{equation}
%%and $\Omega$ the symplectic matrix
%%\begin{equation}
%%\label{OB}
%% \Omega =  \left(
%%\begin{array}{cc}
%%0 & -1 \\
%%1 & 0
%%\end{array}
%% \right).
%%\end{equation}
%%The critical point of the system is $\vec{r}_c=(1,\xi)$.

In spite of non-linearity of the equations of motion given in
(\ref{O2}) an analytical solution is possible, namely,
\begin{equation}
q \;p^{\xi} =  e^{R+q+p}, \label{OVU}
\end{equation}
where R is an integration constant. This takes us to define
\begin{equation}
 \mathcal{M}=\xi \ln p-p+\ln q-q ,
\label{M}
\end{equation}
since $\mathcal{M}=R$ is a constant of motion. Moreover, it can be
rapidly checked that the insertion of (\ref{M}) into Eqs.
(\ref{MEM}) yields the LV equations given in (\ref{O2})
\cite{notaKerner}.  Thus LV is a $\mathcal{M}$ dynamical system.
%Note that the q and p terms in Eq.
%(\ref{M}) can be grouped in the form $\mathcal{M}=V(q)+T(p)$. During
%the evolution there is a flow of mutual resource between $V$ and
%$T$, keeping their sum constant in time.

%Its resource function given in (\ref{M}) can be interpreted as as a
%measure of a mutual ecological resource feeding the system ({\bf or
%better as an entropy?}).

%%In fact, let us define the functions
%%\begin{eqnarray}
%% T&=&\xi \ln p -p,
%% \nonumber\\
%% V&=&\ln q-q,
%%\label{TV}
%%\end{eqnarray}
%%so $\mathcal{M}(q,p)=T(p)+V(q)$. These three functions are plotted
%%in Fig. \ref{figTV}. As can be seen in the graph, the mutual
%%resource $R$ keep constant in the evolution while being distributed
%%between $T$ and $V$.
%%\begin{figure}
%%\includegraphics[scale=0.8,keepaspectratio=true]{plot2_3cycles}
%%\caption{\label{figTV} Time evolution of $\mathcal{M}$ and the T,V
%%functions given in Eqs. (\ref{TV}). The mutual resource $R=T+V$
%%redistributes between the $q$ and $p$ channels.}
%%\end{figure}

%\subsection{Self-feeding systems}
This ecological interpretation of the  $\mathcal{M}$-function as a
mutual abstract resource shared by the interacting species is
enforced by considering a simple self-feeding system, modeled by the
equations of motion
\begin{eqnarray}
\dot{q}&=& q p,
\nonumber\\
\dot{p} &=& -qp . \label{self-feeding}
\end{eqnarray}
%%The self-feeding system is a simplified example of a predator-prey
%%system where the linear death and birth rates of each populations
%%can be neglected in the time scale of consideration. For instance,
consider direct disease transmission \cite{Vandermeer2003} in a
total population of $q+p$ individuals where q are the number of
infected individuals and p the number of individuals who are
susceptible of infection.
%Here the time of spread of the non-lethal
%disease is much shorter than the lifetime of the population, so
%there is no need for inclusion of birth or death rates of the
%carriers.
%The system has analytical solution, namely
%\begin{eqnarray}
%q(t)&=& q_o \frac{R e^{Rt}}{p_o+q_o\, e^{Rt}} ,
%\nonumber\\
%p(t)&=& p_o \frac{R}{p_o+q_o\, e^{Rt}}, \label{castillo}
%\end{eqnarray}
%where $R=q_o+p_o=\mathcal{M}$. The above solutions are represented
%in Fig. \ref{fig3}. As can be seen in the plot, the resource
%function $\mathcal{M}$ is a constant of movement. The phase space of
%the system is given in Fig. \ref{figSFPS}. As can be seen in the
%plot, the number of infected individuals $q$ grows from the initial
%value $q_o$ to the asymptotic limit $R$. In correspondence, the
%number of uninfected individuals decays linearly to zero.
In this simple system there is an  obvious conserved quantity: the
total population given by the initial condition $R=q_o+p_o$. $R$ can
be interpreted as a mutual global resource being distributed between
the "interacting species" $q$ and $p$ that remains conserved during
the process.
%({\bf or as a entropy that scales linearly with the
%number of indiviuals?})
%$R$ also sets the time of spread of the infection
%$R^{-1}$. The bigger the population the quicker the disease is
%transmitted.
The corresponding $\mathcal{M}$-function is then
\begin{equation}
\mathcal{M}=q+p. \label{Msf}
\end{equation}
In fact, it can be checked rapidly that the equations of motion
(\ref{self-feeding}) are obtained by inserting (\ref{Msf}) into Eqs.
(\ref{MEM}).

%\begin{figure}
%\includegraphics{plot3_disease}
%\caption{\label{fig3} Solutions for $q$ (thin solid line) and $p$
%(dashed line)  in a self-feeding system as a function of $\tau=Rt$.
%The resource function $\mathcal{M}$ (thick solid line) remain
%constant in the evolution.}
%\end{figure}

%\begin{figure}
%\includegraphics[scale=0.8]{SelfFeedingPhaseSpace}
%\caption{\label{figSFPS} Phase space for the self-feeding system
%(solid line) for a initial value $q_o$ of 20\% of infected
%individuals. The resource function $\mathcal{M}=q+p$ is also
%represented as a dashed line.}
%\end{figure}

%%We also note that the principle $\delta \mathcal{P}=0$ is also
%%satisfied in this system, since $p=q/\dot{q}$ satisfies Eq.
%%(\ref{dS}).

%\subsection{Logistic evolution}

Consider now the $\mathcal{M}$-function for a two-species Malthusian
system
\begin{equation}
\mathcal{M}=-\gamma \ln q  +k \ln p,
 \label{Mlogistic0}
\end{equation}
so $q=q_o e^{k t}$, $p=p_o e^{\gamma t}$. Consider now the coupling
\begin{equation}
\mathcal{M}=-\gamma \ln q +k \ln p -\mu q \ln \frac{p}{p_s} ,
 \label{Mlogistic}
\end{equation}
where $p_s\ge p_o$ is a saturation constant. The canonical equations
of motion given in Eqs. (\ref{MEM}) equipped with (\ref{Mlogistic})
yield
\begin{eqnarray}
\dot{q}&=& k q -\mu q^2,
\label{m-logistic1}\\
\dot{p} &=& \gamma p +\mu \;q p  \ln \left(p/p_s\right) .
\label{m-logistic}
\end{eqnarray}

Eq. (\ref{m-logistic1}) defines logistic evolution, where $k$ is the
growing rate for the population $q$. The mortality rate $\mu q$ is
density dependent, making the logistic evolution a simple model to
describe the phenomena of saturation of population growth in time
\cite{Krebs,Poole}.
%nonlinear phenomena in population biology \cite{Krebs}. Eq.
%(\ref{m-logistic1}) is a simple form

%Its
%analytical solution is given by
%\begin{equation}
%q(t)=\frac{q_s}{1+e^{-kt}[\frac{q_s}{q_o}-1]},
% \label{logisticsolution}
%\end{equation}
%where  $q_o$ is the initial population. This s-shape solution
%describes an initial exponential increase characterized by the
%growth rate $k$ saturating to  the value $q_s=k/\mu$ at a typical
%time scale given by $k$ (see Fig. \ref{fig4}).

Now we turn to the analysis of Eq. (\ref{m-logistic}). Remarkably,
Eqs. (\ref{m-logistic1}) and (\ref{m-logistic}) have both analytical
solutions. $p$ is the amount of food supply feeding the system, and
thus being depleted by $q$. We take $\gamma=0$ and the change of
variables $\tau=k t$, $\bar{p}=p/p_s$ and $\bar{q}=q/q_s$ where
$q_s$ is the saturation value $q_s=k/\mu$. Eqs.
(\ref{m-logistic1})-(\ref{m-logistic}) reads now
\begin{eqnarray}
\frac{d\bar{q}}{d\tau}&=& \bar{q} (1-\bar{q}),
\label{m-logistic3}\\
\frac{d \bar{p}}{d\tau} &=& \;\bar{q} \bar{p}  \ln \bar{p} .
\label{m-logistic4}
\end{eqnarray}

An analytical solution of Eqs.
(\ref{m-logistic3})-(\ref{m-logistic4}) can be found in the form
\begin{eqnarray}
\bar{q}&=& \frac{e^{\tau}}{e^{\tau}-1+\frac{1}{\bar{q}_o}},
\label{solq}\\
\bar{p}&=& \exp{\left[\alpha \left(
e^{\tau}-1+\frac{1}{\bar{q}_o}\right)\right]}, \label{solp}
\end{eqnarray}
where $\alpha\leq 0$ is an integration constant fixing
$\bar{p}_o=e^{\alpha/\bar{q}_o}$. Therefore, $\alpha$ depends on the
distance between $p_o$ and $p_s$. It is easy to check that the
solutions obtained make $\mathcal{M}$ constant in time.
%In fact, inserting
%(\ref{logisticsolution}) and (\ref{solp}) into Eq. (\ref{Mlogistic})
%we obtain
%\begin{equation}
%\mathcal{M}(q,p)= R+\alpha k
%\left(\frac{q_s}{q_o}-1\right)=\mathcal{M}(q_o,p_o).
% \label{Mlogistic2}
%\end{equation}

\begin{figure}
\includegraphics[scale=0.8]{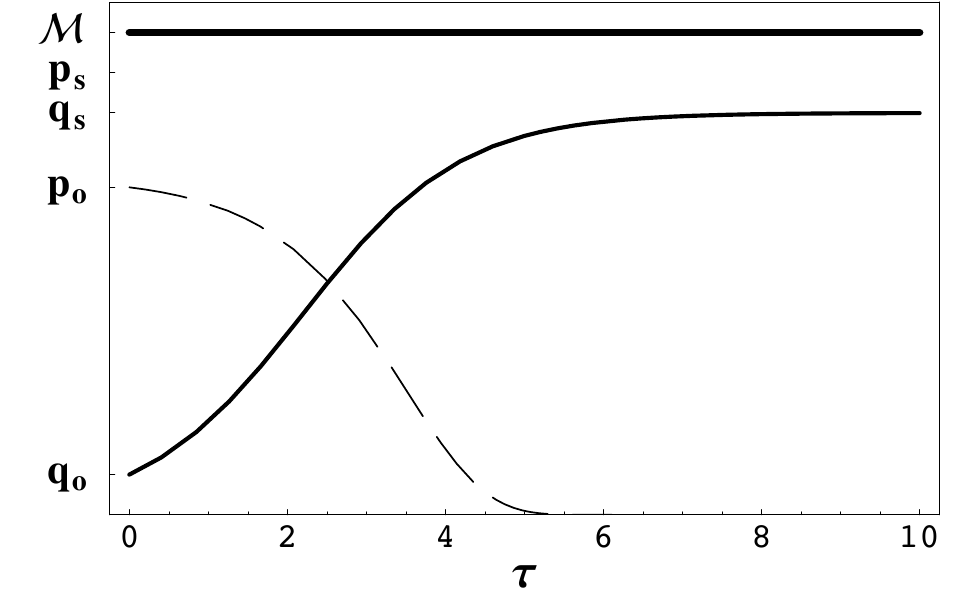}
\caption{\label{fig6} Logistic solutions for $q$ (thin solid line)
and $p$ (dashed line)  as a function of $\tau=k t$ for
$\alpha=-0.03$, $q_o= 0.14$, $q_s= 1.41$, $p_s= 1.55$ and $k=10$.
%Here  $k=\mu$ so the saturation value
%$q_s=1$ coincides with p.
The resource function (thick solid line) remains constant in the
evolution $\mathcal{M}=k(\ln p_s -D)$.}
\end{figure}

The solution (\ref{solp}) has two stationary limits, $p=p_s$ and the
asymptotic limit  $p=0$, since both limits make the right hand side
of Eq. (\ref{m-logistic4}) vanish. They correspond to $\alpha=0$ and
($\alpha<0, \tau\rightarrow\infty$), respectively.
%The logistic growth have been measured in stable environments with a
%limited space feeded externally at a constant rate in order to
%warrant constancy of the food supply in time. The undepleted
%solution $p=p_s$ correspond to this situation.
%%This is shown in Fig. \ref{fig4} where
%%Eqs. (\ref{solp}) and (\ref{Mlogistic2}) are plotted for $\alpha=0$
%%together with Eq. (\ref{logisticsolution}).
%Solutions for $\alpha<0$ models another experimental scenario where
%depletion of the food supply is allowed.
The parameter $\alpha$ measures %the distance from $p_o$ to $p_s$. It characterizes
how fast the food supply is depleted; giving the width of the decay
as $\Delta=\ln (1-1/\alpha)$. This is shown in Fig. \ref{fig6} where
Eqs. (\ref{solq}) and (\ref{solp}) are plotted for $\alpha=-0.03$.
%%The width of the exponential decay have a slow
%%variation with $\alpha$, namely $\Delta \tau=\ln (1-1/\alpha)$.
%Also note that now
%the value of $\mathcal{M}$ is lesser than before due to the
%nonvanishing second contribution in Eq. (\ref{Mlogistic2}).

An explicit form of $\bar{p}$ as a function of $\bar{q}$ can be
obtained after elimination of the time variable between Eqs.
(\ref{solq}) and (\ref{solp}), namely
\begin{equation}
\bar{p}=e^{-D/(1-\bar{q})}, \label{pepino}
\end{equation}
where $D=\alpha (1-1/\bar{q}_o)$. The above equation provide us of
the phase space of the system.
%as plotted in Fig. \ref{fig7}.

%Increasing the value of $R$ leads to a quicker decay of $p$ versus
%$q$. This can be seen in Fig. \ref{fig7} where the phase space of
%the system associated to Fig. \ref{fig6} ($R=1.1$) and for $R=2$ is
%shown.

%\begin{figure}
%\includegraphics{plot4_logistic}
%\caption{\label{fig4} Logistic solutions for $q$ (thin solid line)
%and $p$ (dashed line)  as a function of $\tau=k t$ for $\alpha=0$,
%$q_o= q_s/10$ and $R=1.1$. Here  $k=\mu$ so the saturation value
%$q_s=1$ coincides with p. The resource function (thick solid line)
%remains constant in the evolution, i.e., $\mathcal{M}=R$ .}
%\end{figure}

%\begin{figure}
%\includegraphics[scale=0.8]{LogisticPhaseSpace}
%\caption{\label{fig7} Logistic phase space associated to Fig.
%\ref{fig6}. }
%\end{figure}

%%%%%%%%%%%%%%%%%%%%%%%%%%%%%%%%%%%%%%%%%%%%%%%%%%%%%%%%%%%%%%%%%%%%%%%%%
%

%\section{Discussion}
%\label{SectionDiscussion}

%However, in some cases there exist a conserved quantity in the
%evolution that can be interpreted as a resource function for the
%system.

%Unfortunately, in its original form Lotka-Volterra has some
%significant problems. It predicts cycling behaviour for any set of
%values of the model's four parameters. While this cycling has been
%observed in nature, it is not overwhelmingly common. It appears that
%Lotka-Volterra by itself is not sufficient to model many
%predator-prey systems. Context specific information must be added.

%Hamilton formalism of classical mechanics have a number of
%advantages over the Lagrangian formulation. The equation of motion
%acquires a simpler form. Actually, the formalism yields equation of
%motions as explicit expressions for the time derivatives over a
%Hamiltonian function.

%
%\section{Conclusions}

%%In population dynamics phenomenological rate equations are ascribed
%%to each particular system in order to model their evolution.
%%Different equations are phenomenologically introduced
%%\cite{Hoppensteadt} to model specific situations.

 %In this paper a phase space theory for population dynamics in ecology
 %%which exploits the parallelism with hamiltonian theory of mechanical systems
%has been presented. It applies to $\mathcal{M}$-systems where a
%conserved quantity can be defined in phase space.

The analogy between $\mathcal{M}$-systems and hamiltonian systems
can be exploited further. In this paper, a geometrical approach has
been developed based in the introduction of a $\mathcal{M}$-bracket
which, as the Poisson bracket of hamiltonian systems, has the
structure of a Lie algebra. The $\mathcal{M}$-bracket establishes a
bridge than can be used to generalize the breadth of analytical
methods in phase space developed for the Poisson bracket
\cite{Saletan} (Noether theorem connecting symmetry and conservation
laws,  Liouville theorem establishing conservation in time of phase
space volume for ensembles, etc).
%Also, stochasticity and spatial effects, ignored in
%the present paper, can be incorporated in the phase space treatment.

%Not every dynamical system in ecology is a $\mathcal{M}$-system. We
%have given necessary and sufficient conditions for this.

%Several important models used in population ecology can be shown to
%be $\mathcal{M}$-systems. In this paper, we prove it for
%Lotka-Volterra predator-prey dynamics, the self-feeding systems use
%to model disease transmission and the logistic evolution.

%Finally,  it has been shown that the equation of motion for
%$\mathcal{M}$-systems can be derived from a variational principle
%over a functional $\mathcal{J}$. In $\mathcal{M}$-systems, the
%actual trajectories are such that make $\mathcal{J}$ an extremal
%($\delta \mathcal{J}=0$).

%
%%%%%%%%%%%%%%%%%%%%%%%%%%%%%%%%%%%%%%%%%%%%%%%%%%%%%%%%%%%%%%%%%%%%%%%%%
%%%%%%%%%%%%%%%%%%%%%%%%%%%%%%%%%%%%%%%%%%%%%%%%%%%%%%%%%%%%%%%%%%%%%%%%%
%

\acknowledgements

The author thanks ecologists Antonio P\'erez and F. Garc\'{\i}a Novo
for pointing out some helpful bibliography.
% and my son J. Mart\'{\i}nez-Manso, true inspirer of this work.
J.M-L is supported by a Return Program from the Consejer\'{i}a de
Educaci\'on y Ciencia de la Junta de Andaluc\'{i}a in Spain.

%%\appendix*
%\appendix
%\section{}
%\label{AppendixI}
%
%In this appendix, one can check by inspection that the M-bracket
%defined in Eq. (\ref{MB}) satisfy the following properties:
%\begin{itemize}
%\item Bilinearity: $[X,\alpha Y+\beta Z]=[X,Y]\alpha+[X,Z]\beta$, $[\alpha X
%+\beta Y ,Z]=\alpha [X,Z]+\beta [Y,Z]$.
%\item Antisymmetry: [X,Y]=-[Y,X].
%\item Jacobi identity: [X,[Y,Z]]+[Y,[Z,X]]+[Z,[X,Y]]=0,
%\end{itemize}
%where X,Y,Z are dynamical variables in phase space. This can be deduced through the linearity of the operator
%$\widetilde{\partial}_i = \frac{\partial}{\partial \ln(\eta_i)}$ and
%the antisymmetric character of the symplectic matrix $\Omega$

%
%%%%%%%%%%%%%%%%%%%%%%%%%%%%%%%%%%%%%%%%%%%%%%%%%%%%%%%%%%%%%%%%%%%%%%%%%
%%%%%%%%%%%%%%%%%%%%%%%%%%%%%%%%%%%%%%%%%%%%%%%%%%%%%%%%%%%%%%%%%%%%%%%%%
%
%   REFERENCES


\begin{references}

\bibitem{Goldstein} H. Goldstein, C.P. Poole and J.L. Safko \emph{Classical Mechanics} 3 edition (Addison Wesley, NY,
2002).

\bibitem{Saletan} J.V. Jose and E.J. Saletan, \emph{Classical Dynamics},
Cambridge University Press (1998).

\bibitem{Begon} M. Begon, J.L. Harper and C.R. Townsend,  \emph{Ecology} (Blackwell Science, Cambridge, 1996).

\bibitem{Hoppensteadt} Frank C. Hoppensteadt, \emph{Mathematical Methods in Population Biology}
(Cambridge University Press, 1982).



\bibitem{Sat86} D.\ H.\ Sattinger y O.\ L.\ Weaver,
{\it Lie Groups and Algebras}, (Springer-Verlag, New York, 1986).


\bibitem{Moyal} Another well known example of $C^*$ Lie algebra is the one associated to the Moyal bracket,
generating time displacements of the Wigner function. This
geometrical approach yields the phase space formalism of Quantum
Mechanics given in  J. Moyal, Proc. Camb. Phil. Soc. \underline{45},
99 (1949).




%\bibitem{commentdimensionM} Note that $\mathcal{L}$, as $\mathcal{M}$, has dimension of
%frequency ($[\mathcal{L}]=T^{-1}$). $\mathcal{M}$ gives the rate in
%time at which a mutual resource is available. Thus, $\mathcal{J}$ is
%a dimensionless quantity.


\bibitem{commenthamiltonprinciple} For instance start form the modified Hamilton principle in phase space, which imposes Euler-Lagrange equations to $\mathcal{L}$, as can be seen in \cite{Saletan}.

\bibitem{commentHMP} This ambiguity is inherent to the Hamilton
modified principle. In fact, it also applies to the definition of
$L$, as can be seen in  \cite{Goldstein}.



\bibitem{Lotka}  Lotka, A. J. 1925. \emph{Elements of physical biology}. Baltimore: Williams and Wilkins
Co. Reissued as \emph{Elements of Mathematical Biology}. Dover, New
York, 1956. pp 88-90.

\bibitem{Volterra} Volterra, V. 1926.
%\emph{Variazioni e fluttuazioni del numero
%d'individui in specie animali conviventi}.
Mem. R. Accad. Naz. dei Lincei. Ser. VI, vol. 2. A complete
translation
%with title \emph{Variations and Fluctuations of popular
%size in coexisting animal species}
can be found in Applicable Mathematics of Non-physical Phenomena, F.
Oliveira-Pinto and B.W. Conolly. John Wiley and Sons. 1982. pp
23-115.


\bibitem{Hsu}  S.-B. Hsu, %\emph{A remark on the period of the periodic solution in the Lotka-Volterra system}.
J. Math. Anal. Appl. {\bf 95}, 428-436 (1983).


\bibitem{notaKerner}This fact was already found in the old papers \cite{Kerner} for the particular LV
system. However, our algebraical approach allows us to construct a
phase space formalism for general $\mathcal{M}$-systems.

\bibitem{Kerner} E.H. Kerner, Bull. Math.  Biophys., 19, 121-146
(1957), E.H. Kerner, Bull. Math. Biophys., 21, 217-255 (1959), E.G.
Leigh Jr., Zoology, 53, 777-783 (1965).




\bibitem{Vandermeer2003} J.H. Vandermeer and D. E. Goldberg \emph{Population Biology}.
(Princeton University Press, New Jersey, 2003).

\bibitem{Krebs} C.J. Krebs, \emph{Ecology: the experimental analysis of
distribution of abundance}. Harper and Row (1985).

\bibitem{Poole} For a list of assumptions implicit in the LV and logistic
equations see for instance R.W. Poole, \emph{Quantitative Ecology},
(McGraw-Hill Series in Population Ecology, NY, 1974), pgs 148 and 63
respectively.


\end{references}
\end{document}